\documentclass{emulateapj}

\newcommand{\be}{\begin{enumerate}}
\newcommand{\ee}{\end{enumerate}}

\newcommand{\logoh}{$\log(\mbox{O}/\mbox{H})$}
\newcommand{\tlogoh}{$12+\log(\mbox{O}/\mbox{H})$}

\usepackage{verbatim}

\shorttitle{Massive Metal-Poor Galaxies}
\shortauthors{Peeples, Pogge, \& Stanek}

\date{\today}

\begin{document}

\title{Outliers from the Mass--Metallicity Relation II:\\
 A Sample of  Massive Metal-Poor Galaxies from SDSS}

\author{Molly S.\ Peeples, Richard W.\ Pogge, \& K.\ Z.\ Stanek}
    \affil{Department of Astronomy, Ohio State University, 140 W.\ 18th
    Ave., Columbus,~OH~43210,\\ molly@astronomy.ohio-state.edu, pogge@astronomy.ohio-state.edu, kstanek@astronomy.ohio-state.edu}

\begin{abstract}
We present a sample of 42 high-mass low-metallicity outliers from the
mass--metallicity relation of star-forming galaxies.  These galaxies
have stellar masses that span $\log(M_{\star}/\mbox{M}_{\odot}) \sim
9.4$ to 11.1 and are offset from the mass--metallicity relation by
$-0.3$ to $-0.85$\,dex in \tlogoh.  In general, they are extremely blue,
have high star formation rates for their masses, and are morphologically
disturbed.  Tidal interactions are expected to induce large-scale gas
inflow to the galaxies' central regions, and we find that these
galaxies' gas-phase oxygen abundances are consistent with large
quantities of low-metallicity gas from large galactocentric radii
diluting the central metal-rich gas.  We conclude with implications for
deducing gas-phase metallicities of individual galaxies based solely on
their luminosities, specifically in the case of long gamma-ray burst
host galaxies.
\end{abstract}


\section{Introduction}\label{sec:intro}

Star-forming galaxies fall on a luminosity--metallicity relation such
that more luminous galaxies tend to have higher gas-phase metallicities
(the proxy for which is typically the oxygen abundance in units of
{$12+\log[\mbox{O}/\mbox{H}]$).  This relation is observed to
hold---shifted to lower metallicities---at redshifts as high as $z\sim
3.5$ \citep{erb06a,maiolino08}, and its scatter decreases to $\sim
0.15$\,dex at $z\sim 0$ when galaxy stellar mass replaces luminosity in
the relation \citep{tremonti04}.  Most commonly-accepted theories as to
the origin of the mass--metallicity relation have as a central
proposition that low-mass galaxies are metal-deficient, rather than that
high-mass galaxies are metal-enhanced \citep{larson74,
dalcanton07,finlator08}.  In fact, there is some evidence that the
mass--metallicity relation may flatten at large stellar masses,
$\log(M_{\star}/\mbox{M}_{\odot}) \gtrsim 10.5$
\citep[e.g.,][]{tremonti04}, though it is unclear to what extent this
flattening is due to a saturation of the metallicity indicator used at
high oxygen abundances \citep[see
e.g.,][]{kewley02,bresolin06,kewley08}.

Because most star-forming galaxies {\em do} lie on a mass--metallicity
locus, we can also learn about the gas-phase metallicity evolution of
galaxies by studying the properties of galaxies that do {\em not} fall
on the relation.  In \citet[][hereafter Paper~I]{peeples08}, we explored
the population of low-mass high-metallicity outliers, and postulated
that these metal-rich dwarf galaxies have low gas fractions and are
therefore nearing the end of substantial epochs of star formation.  Here
we investigate the other corner of the mass--metallicity plane by asking
what we can learn about the evolution of massive galaxies from the
properties of the high-mass low-metallicity outliers.

As shown in Figure~\ref{fig:ohmstar}, we find 42 low-metallicity
high-mass galaxies with masses ranging from
$\log(M_{\star}/\mbox{M}_{\odot}) \sim 9.4$ to 11.1 and offsets from the
central mass--metallicity relation of $-0.3$ to $-0.85$\,dex.  We
describe in \S\,\ref{sec:sample} how we selected this sample and
verified the galaxies' outlier status.  In \S\,\ref{sec:disc}, we
describe the physical properties of these galaxies and discuss possible
origins for their low oxygen abundances.  Specifically, we find that
they have highly disturbed morphologies strongly suggestive of merging
or post-merging systems, are extremely blue, and have high specific star
formation rates.  We summarize our conclusions and state some
implications of these findings in \S\,\ref{sec:conc}.

\begin{figure*}
\plotone{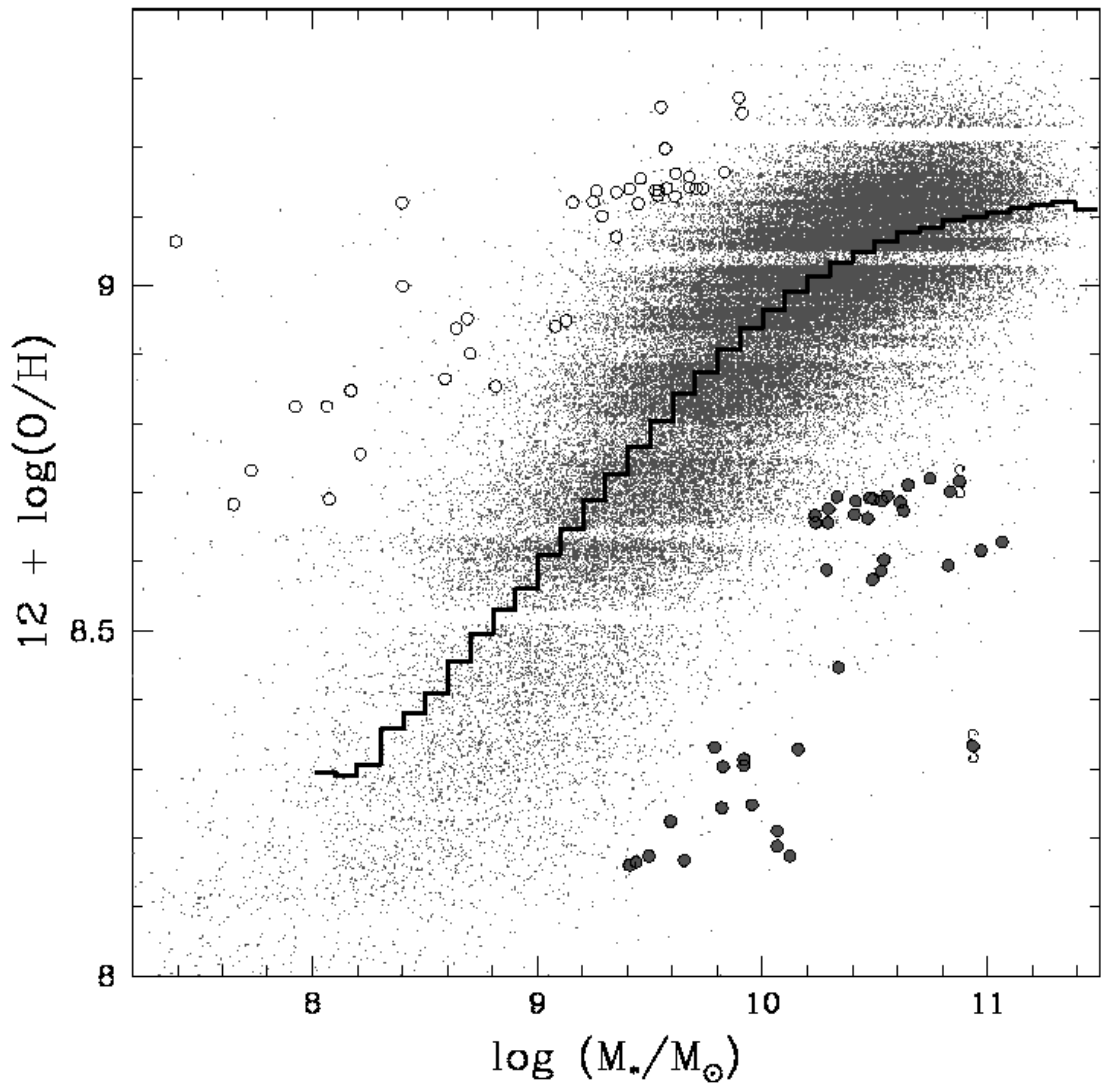}
\caption{\label{fig:ohmstar}Plot of \tlogoh\ v.\
  $\log(\mbox{M}_{\star}/\mbox{M}_{\odot})$. The small grey points
  represent galaxies from the main \citet{tremonti04} SDSS sample
  described in \S\,\ref{sec:sample}, and the step-curve is the median of
  these points in bins of $0.1$\,dex of total stellar mass.  The
  high-mass low-metallicity massive galaxies in our sample are plotted
  as the large grey circles in the lower-right region of the diagram;
  the only two clearly spiral galaxies in the sample are denoted with
  ``\S'' symbols.  For reference we also plot the high-metallicity
  outliers from Paper~I as large open points ({\em upper left}).  }
\end{figure*}

\section{Selecting High-Mass Low-Metallicity Galaxies}\label{sec:sample}
We began with a sample of $\sim 110000$ star-forming galaxies from the
Sloan Digital Sky Survey (SDSS) Data Release 4 \citep{adelman06} with
gas-phase abundances measured by \citet{tremonti04} and stellar masses
derived using the techniques of \citet{kauffmann03}. High-luminosity
outliers from the mass--metallicity relation can be found in the
low-metallicity region of parameter space for one of two reasons: either
they have spuriously low derived metallicities or they are genuine
outliers.  (The third possibility of an object having a spuriously high
measured luminosity is much less likely.)  In general, the strengths of
the visible-wavelength oxygen forbidden [\ion{O}{2}] and [\ion{O}{3}]
emission lines increase as the oxygen abundance decreases; hence, it is
possible for a bright galaxy with a weak active galactic nucleus (AGN)
component to appear to have a low oxygen abundance due to contribution
from the AGN itself.  Likewise, an otherwise ``red and dead'' galaxy
with strong [\ion{O}{2}] emission might be labeled as star-forming and,
in the Bayesian analysis of \citeauthor{tremonti04}, be assigned an
artificially low metallicity.  We therefore selected our sample of
massive low-metallicity galaxies by first ensuring that they are
outliers in the luminosity-- and mass--metallicity parameter spaces and
then following up with several line-ratio diagnostic tests in an attempt
to guarantee that their oxygen abundances relative to the rest of the
sample are indeed believable.  Images of the galaxies in our final
sample are shown in Figure~\ref{fig:images} and summary information is
provided in Table~\ref{tbl:sample}.  As only two of the galaxies in
Figure~\ref{fig:images} are clearly undisturbed spirals, we note these
two galaxies in Table~\ref{tbl:sample}, Figure~\ref{fig:ohmstar}, and
subsequent figures with ``\S'' symbols.

\begin{figure*}
\plotone{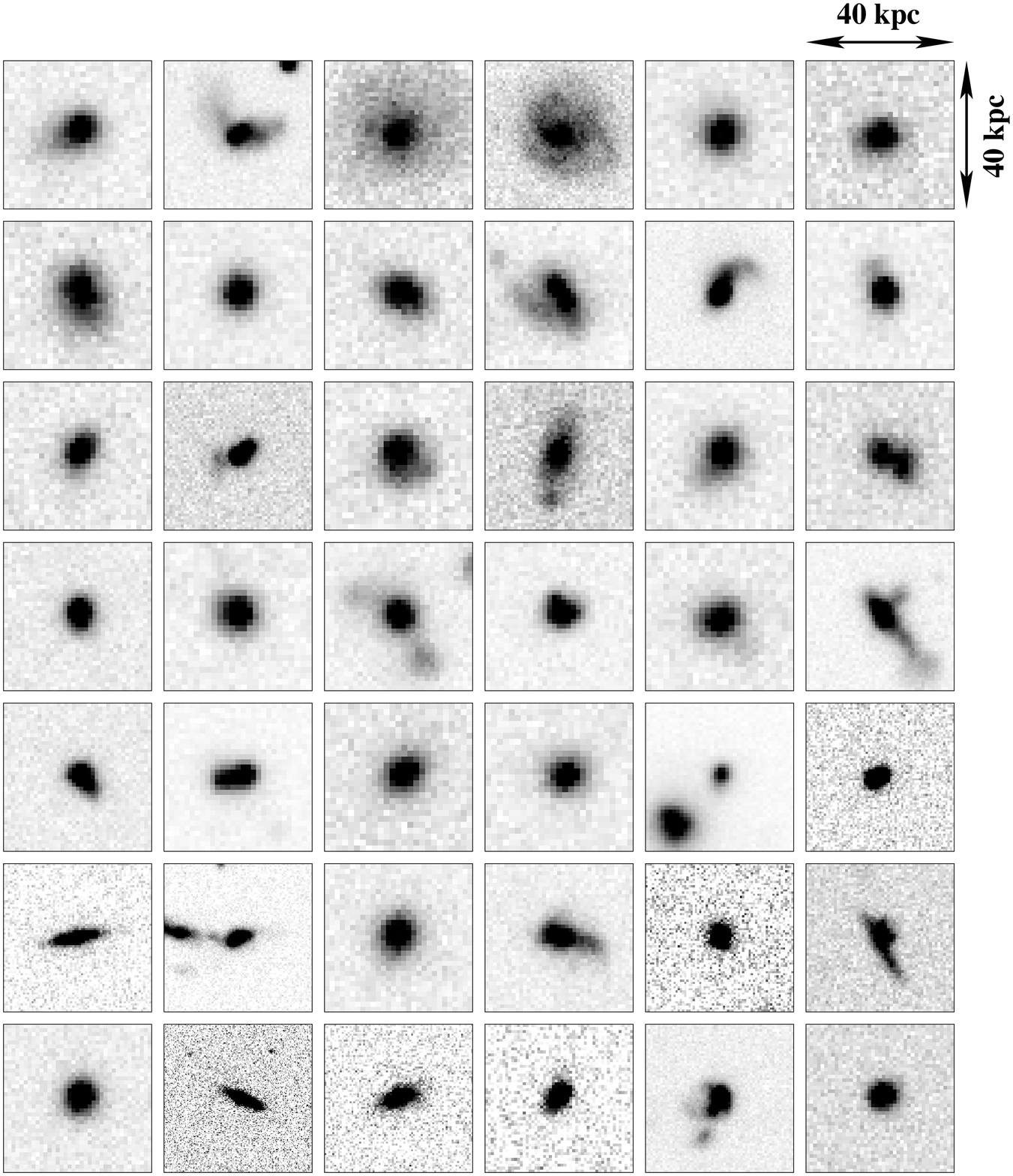}
\caption{\label{fig:images}Normalized SDSS $g$-band images of massive
  low-metallicity galaxies scaled to $40\times 40$\,kpc, with mass
  decreasing to the right and down from $\log M_{\star}\approx\,
  11.1$ to 9.4. The only two clearly spiral galaxies are the middle two
  on the top row.}
\end{figure*}

\begin{deluxetable}{rrcrc}
\tabletypesize{\scriptsize}
\tablewidth{0.5\textwidth}
\tablecaption{Low-metallicity mass--metallicity outliers
\label{tbl:sample} }
\tablecolumns{5}
\tablehead{
\colhead{RA} &
\colhead{dec} &
\colhead{\tlogoh} &
\colhead{$\log \mbox{M}_{\star}$} &
\colhead{redshift}}
\startdata
$149.9796$ & $10.24727$ & $8.63$ & $11.07$ & $0.2423$ \\ 
$255.5060$ & $60.79618$ & $8.62$ & $10.97$ & $0.1257$ \\ 
$191.3246$ & $4.87648$ & $8.33$ & $10.94$ & $0.1800$\tablenotemark{\S} \\ 
$168.6324$ & $48.91903$ & $8.72$ & $10.88$ & $0.1116$\tablenotemark{\S} \\ 
$197.4139$ & $62.76859$ & $8.70$ & $10.84$ & $0.2587$ \\ 
$329.9707$ & $1.03852$ & $8.59$ & $10.83$ & $0.2200$ \\ 
$150.2157$ & $8.30726$ & $8.72$ & $10.75$ & $0.2290$ \\ 
$148.0473$ & $54.31252$ & $8.71$ & $10.65$ & $0.2544$ \\ 
$146.8469$ & $53.06987$ & $8.67$ & $10.63$ & $0.2471$ \\ 
$214.2315$ & $40.44983$ & $8.69$ & $10.61$ & $0.2060$ \\ 
$196.1901$ & $62.40580$ & $8.70$ & $10.56$ & $0.1118$ \\ 
$206.4954$ & $11.47993$ & $8.60$ & $10.54$ & $0.2373$ \\ 
$128.9575$ & $44.87386$ & $8.69$ & $10.53$ & $0.2264$ \\ 
$158.5174$ & $6.20286$ & $8.59$ & $10.53$ & $0.1043$ \\ 
$140.1770$ & $0.84295$ & $8.69$ & $10.49$ & $0.2521$ \\ 
$118.9094$ & $33.44449$ & $8.57$ & $10.49$ & $0.1413$ \\ 
$170.4012$ & $0.54706$ & $8.69$ & $10.48$ & $0.2292$ \\ 
$17.1399$ & $-0.46871$ & $8.66$ & $10.47$ & $0.2270$ \\ 
$158.5274$ & $5.51848$ & $8.69$ & $10.41$ & $0.1708$ \\ 
$181.9754$ & $12.39080$ & $8.67$ & $10.41$ & $0.2613$ \\ 
$200.4299$ & $43.40588$ & $8.45$ & $10.34$ & $0.1950$ \\ 
$126.2741$ & $7.21643$ & $8.69$ & $10.33$ & $0.1974$ \\ 
$349.5542$ & $-0.69060$ & $8.68$ & $10.29$ & $0.2517$ \\ 
$218.7960$ & $44.18318$ & $8.66$ & $10.29$ & $0.1276$ \\ 
$250.6480$ & $42.39715$ & $8.59$ & $10.29$ & $0.1511$ \\ 
$244.4054$ & $35.82085$ & $8.66$ & $10.24$ & $0.2259$ \\ 
$196.2852$ & $53.19373$ & $8.67$ & $10.23$ & $0.2719$ \\ 
$146.2970$ & $42.64472$ & $8.33$ & $10.16$ & $0.2576$ \\ 
$240.8967$ & $31.83226$ & $8.17$ & $10.12$ & $0.1544$ \\ 
$127.9271$ & $51.42651$ & $8.21$ & $10.07$ & $0.0813$ \\ 
$153.3421$ & $2.58265$ & $8.19$ & $10.07$ & $0.0780$ \\ 
$48.8150$ & $-7.76643$ & $8.25$ & $9.95$ & $0.0612$ \\ 
$129.3898$ & $47.96454$ & $8.31$ & $9.92$ & $0.2152$ \\ 
$234.2552$ & $31.44251$ & $8.30$ & $9.92$ & $0.1538$ \\ 
$233.5084$ & $-1.96201$ & $8.30$ & $9.83$ & $0.0787$ \\ 
$ 1.2987$ & $1.06057$ & $8.24$ & $9.82$ & $0.1028$ \\ 
$156.5582$ & $48.74970$ & $8.33$ & $9.79$ & $0.1604$ \\ 
$226.0778$ & $2.76445$ & $8.17$ & $9.65$ & $0.0350$ \\ 
$345.7720$ & $1.24968$ & $8.22$ & $9.59$ & $0.0690$ \\ 
$168.1920$ & $1.33819$ & $8.17$ & $9.49$ & $0.1088$ \\ 
$219.6623$ & $53.14538$ & $8.16$ & $9.44$ & $0.0900$ \\ 
$197.1532$ & $59.88513$ & $8.16$ & $9.41$ & $0.1410$ \\ 
\enddata

\tablecomments{ Sample of massive low-metallicity galaxies, reverse
  sorted by stellar mass.  RA and Dec are in degrees (J2000.0), \tlogoh\
  and stellar masses are from \citet{tremonti04}.}

\tablenotetext{\S}{Spiral galaxy.}
\end{deluxetable}

\subsection{Sample Selection}\label{sec:sel}
We summarize the selection process for our massive metal-poor galaxy
sample in Table~\ref{tbl:cuts}.  We begin by defining the parent galaxy
sample from which our galaxies are outliers.  Starting with the entire
\citet{tremonti04} sample of star-forming galaxies, we choose only those
galaxies with SDSS magnitude errors in the $g$, $r$, $i$, and $z$-bands
of $<0.1$\,mag.  In the $u$-band we take those galaxies that satisfy the
conditions
\begin{eqnarray}
\sigma(u) < 0.1 & \mbox{for} & u < 18, \\
\sigma(u) < \left(\frac{0.1}{1.5}\right)u + 16.5 & \mbox{for} & 18 \leq u < 19.5,\, \mbox{and} \\
\sigma(u) < 0.2 & \mbox{for} & u \geq 19.5,
\end{eqnarray}
where $u$ is the observed $u$-band magnitude and $\sigma(u)$ is the SDSS
error on $u$.  These cuts yield a parent population of 86754 galaxies.
To find the low-metallicity outliers from the luminosity--metallicity
locus, we take bins of absolute $g$-band magnitude $M_g$ of width
0.4\,mag, and each bin we take the 1\% of objects with the smallest
\tlogoh, as demonstrated in Figure~\ref{fig:cuts}.  Because we do not
want to preferentially select blue objects, we similarly take the 1\% of
galaxies with the smallest
\tlogoh\ in bins of $M_z$ of width 0.4\,mag.  Finally, we ensure that
these galaxies are also outliers in the mass--metallicity plane by
taking the 2.5\% smallest \tlogoh\ galaxies in bins of stellar mass of
width $\Delta\log\mbox{M}_{\star}=0.1$\,dex and the 2.5\% largest
$\log\mbox{M}_{\star}$ galaxies in bins of \tlogoh\ of width 0.1\,dex.
These selection criteria yield a sample of 113 candidates for detailed
evaluation and further down-selection of the true low-metallicity
outliers.

\begin{center}
\begin{deluxetable}{ll}
\tablecaption{Cuts for Sample Selection
\label{tbl:cuts} }
\tablecolumns{2}
\tablehead{
\colhead{Cut} &
\colhead{Number Surviving}}
\startdata
Low magnitude errors & 86754 \\
1\% low O/H w.r.t.\ $M_g$ & 855 \\ 
1\% low O/H w.r.t.\ $M_z$ & 598 \\ 
2.5\% low O/H w.r.t. $\log\mbox{M}_{\star}$ & 381 \\ 
2.5\% high $\log\mbox{M}_{\star}$ w.r.t. O/H & 113 \\ 
within 2$\sigma$ of \ion{H}{2} region of [\ion{O}{1}] BPT diagram & 56 \\
OK spectrum reduction & 52 \\
reproducible [\ion{N}{2}]/H$\alpha$ ratio & 44 \\
2.5\% low PP04 (O/H) w.r.t. $\log\mbox{M}_{\star}$ & 42
\enddata
\tablecomments{See \S\,\ref{sec:sample} and Figure~\ref{fig:cuts} for a
  more detailed explanation.
}
\end{deluxetable}
\end{center}

\begin{figure}
\plotone{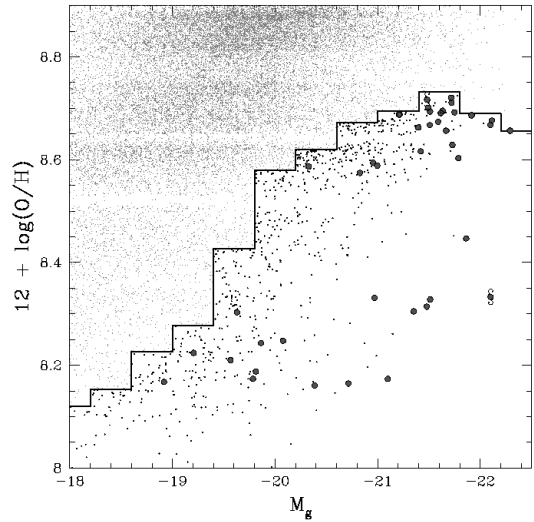}
\caption{\label{fig:cuts} An example of one of the cuts made to find
  low-metallicity high-luminosity outliers from the mass--metallicity
  relation.  Plotted is \tlogoh\ vs.\ $M_g$ for the main sample ({\em
  small light grey points}), the 1\% objects the lowest \logoh\ in bins
  of $M_g$ of width 0.4\,mag ({\em small black points}), and the final
  sample of high-mass low-metallicity outliers ({\em large grey
  points}).  The only two clearly spiral galaxies in the sample are
  denoted with ``\S'' symbols.  The histogram shows the 1\% threshold in
  \tlogoh\ in bins of $M_g$.}
\end{figure}

One possible contamination source of these seemingly bright metal-poor
galaxies is a low-level AGN, such as a low-ionization nuclear
emission-line region (LINER).  While it is possible for galaxies with
weak AGN activity to {\em also} have low oxygen abundances, we choose to
not try to disentangle these two effects on the emission line fluxes.
Using the line fluxes measured by \citet{tremonti04}\footnote{See
\texttt{http://www.mpa-garching.mpg.de/SDSS/DR4/raw\_data.html} for this
very rich dataset.}, we put these galaxies on the standard
\citeauthor*{baldwin81} (BPT) \citeyear{baldwin81} diagrams.  In
particular, as shown in Figure\,\ref{fig:bptn2o1}, we find that about
half of these outliers have strong enough [\ion{O}{1}]$\lambda\,6300$ to
place them in the ``AGN'' region of the BPT diagram.  (We note that
while the \citet{tremonti04} galaxies were chosen to be star-forming
based on where they fall in the [\ion{O}{3}]/H$\beta$ v.\
[\ion{N}{2}]/H$\alpha$ plane, 3.4\% of the parent sample galaxies fall
in the AGN region of the [\ion{O}{3}]/H$\beta$ v.\
[\ion{O}{1}]/H$\alpha$ plane.)  As expected, the strong-[\ion{O}{1}]
galaxies also fall close to the ``composite''
\ion{H}{2}--AGN boundary as defined by \citet{kewley06b} in the
[\ion{O}{3}]/H$\beta$ v.\ [\ion{N}{2}]/H$\alpha$ plane.  Visual
inspection of the spectra and images of these bogus outliers reveals
that many of them are quite red, with only strong
[\ion{O}{2}]$\lambda\,3727$ and perhaps weak H$\alpha$ emission being
suggestive of ongoing star formation.  Because
[\ion{O}{1}]$\lambda\,6300$ is a relatively weak line and many of the
outliers are clustered near the the \citet{kewley06b} boundary, we kept
galaxies whose [\ion{O}{1}]/H$\alpha$ ratio is within 2-$\sigma$ of the
\ion{H}{2}--AGN boundary and whose spectra passed a visual inspection
test for, e.g., clear H$\beta$ emission.  Representative spectra are
shown in Figure~\ref{fig:spec}.

\begin{figure*}
\includegraphics[height=\textwidth,angle=270]{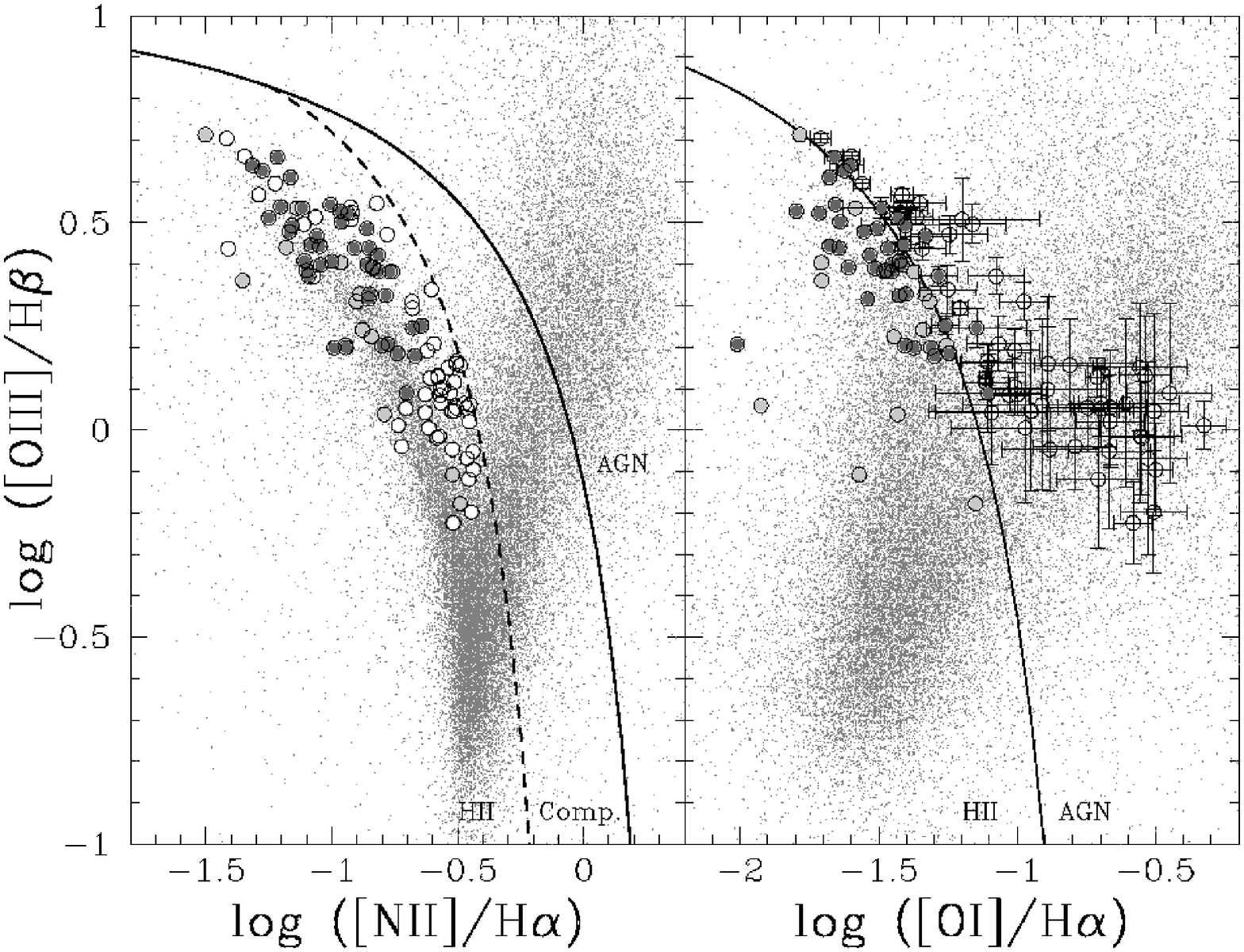}
\caption{\label{fig:bptn2o1}\citeauthor*{baldwin81} diagrams of
  $\log(\mbox{[\ion{O}{3}]}/\mbox{H}\beta)$ v.\
  $\log(\mbox{[\ion{N}{2}]}/\mbox{H}\alpha)$ [{\em left}] and
  $\log(\mbox{[\ion{O}{3}]}/\mbox{H}\beta)$ v.\
  $\log(\mbox{[\ion{O}{1}]}/\mbox{H}\alpha)$ [{\em right}].  Plotted are
  the spectroscopic galaxies from SDSS DR4 ({\em small grey points}),
  the galaxies passing our luminosity, mass, and \logoh\ cuts but not in
  the final sample ({\em open points}), the galaxies passing the BPT
  diagram cuts ({\em light grey points}), and the galaxies in our final
  low-metallicity sample ({\em dark grey points}). Lines are taken from
  \citet{kewley06b}.}
\end{figure*}

\begin{figure*}
\includegraphics[height=\textwidth,angle=270]{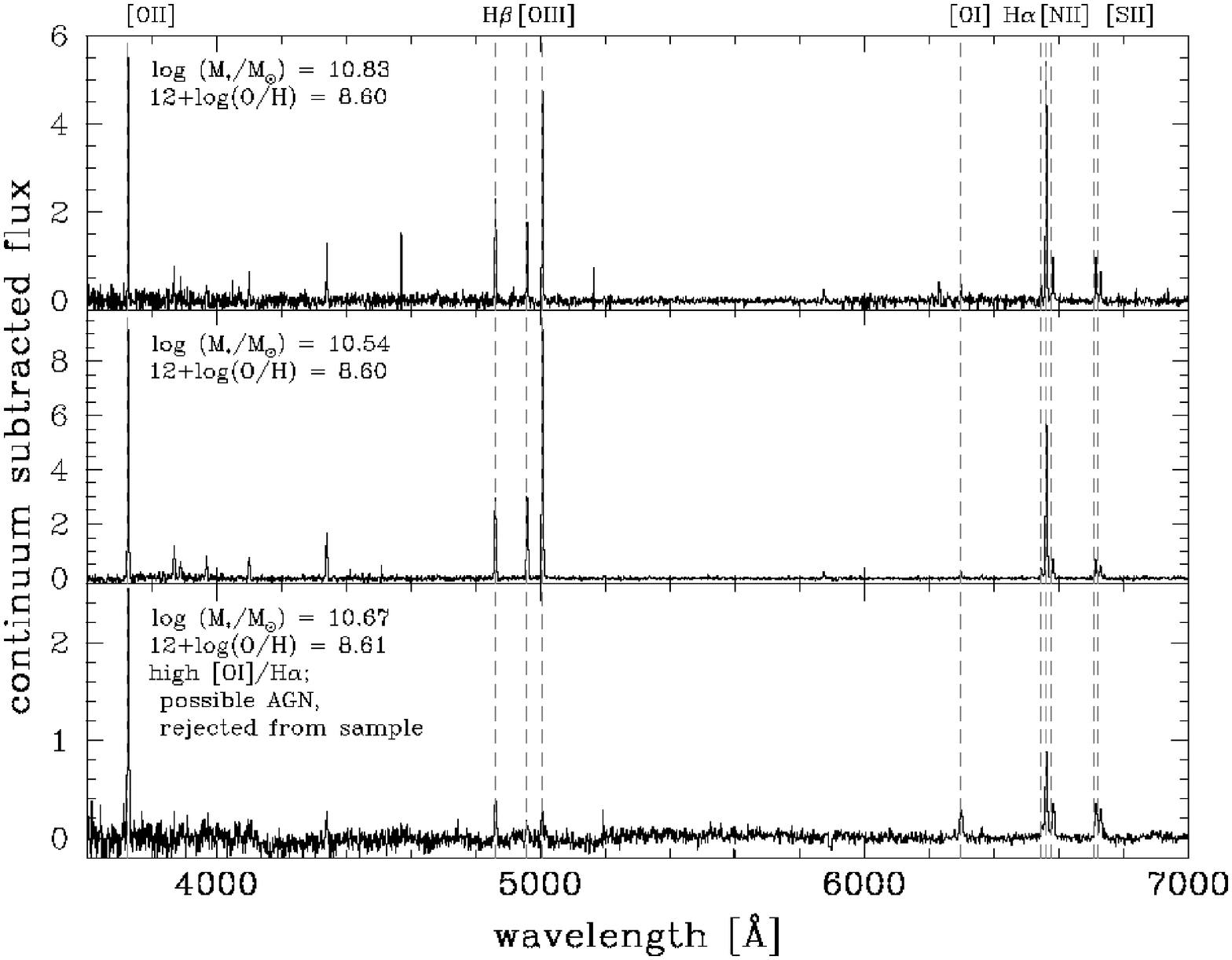}
\caption{\label{fig:spec} Sample spectra.  Two low-metallicity high-mass
outliers ({\em top and middle panels}) and one galaxy rejected as a
potential AGN ({\em bottom}).}
\end{figure*}

We further tested whether or not the measured metallicities could be due
to measurement errors in the relevant line fluxes.  Using the Data
Release 6 SDSS spectra \citep{adelman08}, we subtracted the underlying
stellar continuum using the STARLIGHT program \citep{fernandes05}, which
simultaneously corrects for reddening and extinction using the
\citet{cardelli89} extinction curves.  While the extinction corrections
are all small ($A_V < 1$), the \citeauthor{tremonti04}\ line fluxes do
not account for reddening, and thus a comparison of widely spaced line
ratios produces a systematic offset.  We therefore reject seven galaxies
for which our measurements strongly disagree with
\citeauthor{tremonti04}'s on the [\ion{N}{2}]/H$\alpha$ ratio; the
difference in each of these cases is such that using the
\citeauthor{tremonti04}\ ratio would result in a lower \tlogoh\ than if
ours were used.  The added advantage to using the [\ion{N}{2}]/H$\alpha$
ratio is that this is the ratio used in the \citet{pettini04}
metallicity determination, which we use as discussed below.

\subsection{Metallicity Believability}\label{sec:metals}
We note that we do not impose a metallicity error cut when selecting
these low-metallicity outliers.  This is because the errors on
\citeauthor{tremonti04}'s \tlogoh\ are not a simple function of
metallicity in this range.  It has long been recognized that
accurately measuring low oxygen abundances (e.g., using $R_{23}$) can
be tricky due to degeneracies in the ionization parameter, and the line
ratios usually used to determine the ionization parameter (e.g.,
[\ion{O}{3}]$\lambda$\,5007/[\ion{O}{2}]$\lambda$\,3727) are also
dependent on metallicity \citep{mcgaugh91,kewley02}.  Unfortunately,
this means that many metallicity indicators are discontinuous or
systematically tend to avoid particular abundance solutions.

We therefore adapted a different approach to ensure that our
low-metallicity high-mass outliers geniunely have relatively low oxygen
abundances.  As discussed in \S\,\ref{sec:sample}, we first chose only
those galaxies that lie in the \ion{H}{2} region locus of the BPT
diagrams, and secondly, we discarded those galaxies for which we did not
measure a comparable [\ion{N}{2}]/H$\alpha$ ratio to those given by
\citeauthor{tremonti04}.  Finally, we argue that if these galaxies are
truly high-mass low-metallicity outliers, then they should be outliers
regardless of how \tlogoh\ is measured.  (See \citealt{kewley02} and
\citealt{kewley08} for thorough discussions of how and why different
metallicity callibrations differ.) We therefore used the
\citeauthor{tremonti04}\ measurements recalculate \tlogoh\ for the
entire parent sample using the \citet{pettini04} method, where the
oxygen abundance is given by
\begin{equation}\label{eqn:pp04}
12 + \log(\mbox{O}/\mbox{H}) = 9.37 + 2.03\times\mbox{N2}
                               + 1.26\times\mbox{N2}^2 + 0.32\times\mbox{N2}^3,
\end{equation}
where
$\mbox{N2}\equiv\log([\mbox{\ion{N}{2}}]\lambda\,6584/\mbox{H}\alpha)$.
The main disadvantage of this method is that it uses nitrogen as a proxy
for oxygen; however, there are several factors that make it advantageous
in these objects.  First, it is reddening insensitive because the
[\ion{N}{2}]$\lambda$\,6584 line and H$\alpha$ are separated by only
$\sim 20$\,\AA.  This means that we can safely use the
\citeauthor{tremonti04}\ measurements and our own measurements without
being overly concerned about the reddening corrections.  Second, unlike
the various of R$_{23}$ methods, the \citeauthor{pettini04}\ method is
both continuous and single-valued throughout the metallicity range of
interest; there are no ``upper-'' and ``lower''-branches to concern us.
We show the mass--metallicity relation with \tlogoh\ measured with the
\citet{pettini04} method in Figure~\ref{fig:pp04mstar}.  Only one galaxy
in the low-[\ion{O}{1}] sample did not pass the
\tlogoh--$\log\mbox{M}_{\star}$ cuts described in Table~\ref{tbl:cuts},
suggesting that the remaining galaxies are true low-metallicity
outliers.

\begin{figure}
\plotone{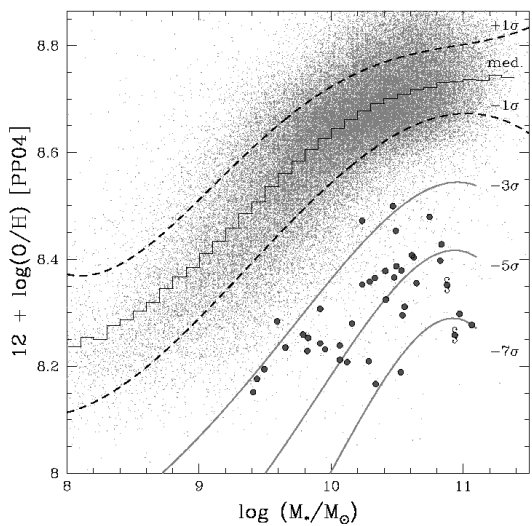}
\caption{\label{fig:pp04mstar}Plot of \citet{pettini04} \tlogoh\ v.\
$\log(\mbox{M}_{\star}/\mbox{M}_{\odot})$. The galaxies in our sample
are plotted as the large grey circles in the lower-right region of the
mass--metallicity diagram; the only two clearly spiral galaxies in the
sample are denoted with ``\S'' symbols. The histogram denotes the median
\tlogoh\ in bins of stellar mass, while the curves are fits to standard
deviation offsets as labeled.}
\end{figure}

We have made several other checks for potential systematics.  First,
spiral (and thus many star-forming) galaxies are known to have
metallicity gradients such that their centers have higher abundances
than at larger radii \citep[e.g.,][]{zaritsky94, kennicutt03}.  It is
therefore reasonable to expect that, on average, spiral galaxies with
larger fractions of their light falling in the 3\arcsec\ SDSS fiber will
have lower integrated metallicities.  However, most of our galaxies (see
Fig.\,\ref{fig:images}) are not spiral galaxies, so we do not expect for
this to be a huge effect.  Regardless, as shown in
Figure~\ref{fig:ohff}, while the galaxies in our sample do have somewhat
high fiber fractions, this is only a small effect, and the offsets we
see in metallicity are much greater than can be explained by large fiber
fractions, which would cause contamination of central high-metallicity
\ion{H}{2} regions by low-metallicity \ion{H}{2} gas at larger
galactocentric radii.

\begin{figure}
\plotone{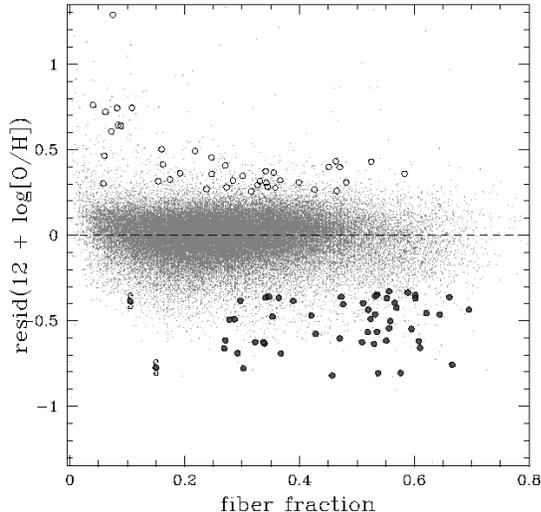}
\caption{\label{fig:ohff}Plot of residual in \tlogoh\ from the
  \citeauthor{tremonti04}\ fit to the mass--metallicity relation v.\
  fiber fraction. The high-mass low-metallicity outliers are shown as
  the large dark grey points at the bottom of the diagram, with 
  the only two distinctly spiral galaxies in the sample as
  ``\S'' symbols.  For reference we plot the high-metallicity outliers
  from Paper~I as large open points ({\em top}).  }
\end{figure}

Similarly, the final redshift distribution for the galaxies in our
sample peaks at higher $z$ than for the main sample; since we have
convinced ourselves that this is not purely a fiber fraction effect, it
is worrisome that it might partially be a redshift-evolution effect.
When the universe was younger, galaxies are expected to have had lower
oxygen abundances; the observed mass--metallicity relations for
redshifts $z>0.1$ are systematically shifted to lower metallicity from
what is observed for nearby galaxies \citep{erb06a,maiolino08}.  We plot
our low-metallicity outliers against the observed mass--metallicity
evolution from \citet{maiolino08} in Figure~\ref{fig:maiolino}.  While
the median redshift for galaxies in our sample is $z \sim 0.17$, the
observed oxygen abundances are more typical of redshifts $2.2 \lesssim z
\lesssim 3.5$.  Though a large amount of scatter in the
mass--metallicity relation is both expected and observed
\citep{savaglio05, erb06a, kobayashi07, finlator08, maiolino08} at all
redshifts, the abundance offsets of our low-metallicity outliers are
much more pronounced than the relatively subtle expectations from
metallicity evolution with cosmic time.

\begin{figure}
\plotone{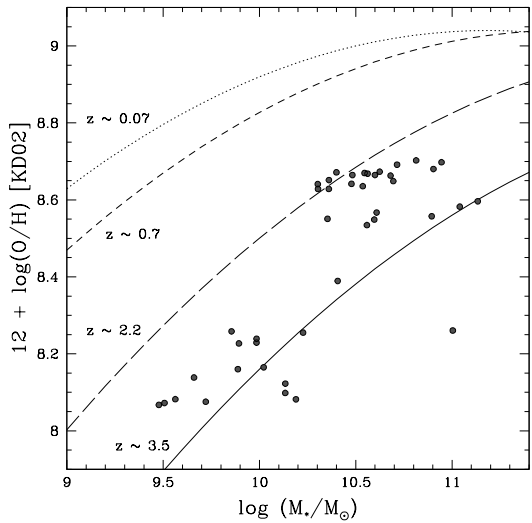}
\caption{\label{fig:maiolino} Evolution of the mass--metallicity
  relation from \citet{maiolino08} with comparison to low-metallicity
  high-mass outliers (grey points).  Fits to the mass--metallicity
  relation from \citet{maiolino08} are shown with lines: $z\sim 0.7$,
  dotted; $z\sim 0.7$, short dashed; $z\sim 2.2$, long dashed; and
  $z\sim 3.5$ solid.  The grey points denote our high-mass
  low-metallicity outliers ($0.035 < z < 0.27$), with re-calculated
  metallicities using the \citet{kewley02} diagnostic using our
  measurements of the line ratios so that they are on the same scale as
  the \citeauthor{maiolino08} results.  (The stellar masses have also
  been shifted to match the \citeauthor{maiolino08} fits.)}
\end{figure}

\section{Discussion}\label{sec:disc}
To seek to explain these galaxies' low oxygen abundances, we need to
consider the population's other physical properties.  As shown in
Figure~\ref{fig:cmd}, the galaxies in our sample are all {\em very}
blue; most---but not all---are also outliers in the galaxy
color--magnitude diagram.  In \S\,\ref{sec:sfr}, we discuss how this
blueness can be attributed to the galaxies' high specific star formation
rates.  Figure~\ref{fig:images} shows that 40 out of 42 of the galaxies
in our sample have disturbed morphologies suggestive of merging systems.
Though some of the less-resolved galaxies in our sample may appear only
slightly morphologically disturbed in the SDSS images, evidence suggests
that, with higher resolution, these galaxies do in fact have unusual
morphologies.  In a {\em Hubble Space Telescope} study of local Lyman
break galaxy (LBG) analogs,
\citet{overzier08} found SDSSJ102613.97+4884458.9 ($[\alpha,\delta] =
[156.5580,48.7497]$, leftmost image on the bottom row of
Figure\,\ref{fig:images}) to have a strongly asymmetric profile with
several distinct knots of star formation.  In a preliminary study of the
most UV luminous galaxies, \citet{hoopes07} found that these high
specific star formation rate LBG analogs are metal-poor by $\sim 0.5$
dex relative to other galaxies of similar stellar mass.
\citeauthor{hoopes07}\ and \citeauthor{overzier08}\ suggest that the
observed high specific star formation rates and low metallicities for
these LBG analogs are related to the galaxy collisions that formed these
objects.  (See also \citealt{struck08} for a discussion of so-called
``delayed'' galaxies, i.e., high star formation rate interacting
galaxies which have apparently managed to retain most of their gas until
$z=0$.)  In \S\,\ref{sec:merge}, we show how a simple merger-induced gas
inflow picture can account for an offset in metallicity, and we discuss
possible shortcomings in this interpretation.  The other two galaxies in
the sample are undisturbed spiral galaxies, and while these are clearly
very interesting objects, we have no explanation to offer for their
extreme offsets from the mass--metallicity relation.

\begin{figure}
\plotone{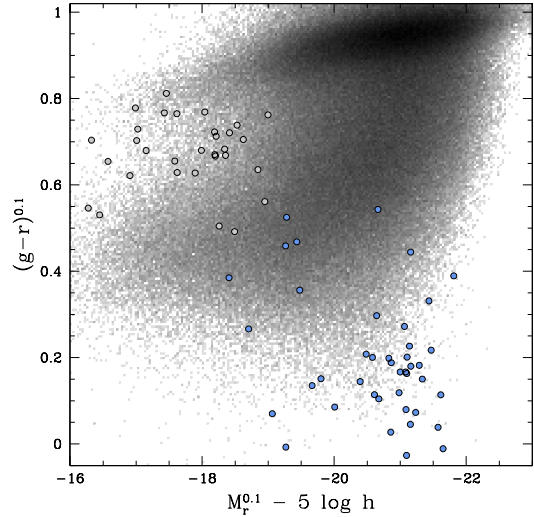}
\caption{\label{fig:cmd} Mass--metallicity outliers on the
  color--magnitude diagram [$(g-r)$ vs.\ $M_r$, both $k$-corrected to
  $z=0.1$ using the templates from \citet{assef08}] from the SDSS DR6
  spectroscopic galaxy sample \citep{adelman08}.  The blue circles
  denote the 42 low-metallicity high-mass galaxies in our sample; for
  reference, metal-rich dwarf galaxies from Paper~I are denoted by the
  grey circles. }
\end{figure}

\subsection{Star Formation Rates}\label{sec:sfr}
\citet{ellison08a} have shown that galaxies with higher specific star
formation rates at a given mass have preferentially lower metallicities.
In Figure~\ref{fig:sfrmm}, we plot the specific star formation rate
(SSFR) as a function of stellar mass for the SDSS star-forming
galaxies. Most (but not all) of our low-metallicity outliers clearly
have higher specific star formation rates than typical of the larger
sample. We plot the specific star formation rates against the
mass--metallicity residual from \citet{tremonti04} in
Figure~\ref{fig:sfroh}; generally speaking, aperture corrections for the
specific star formation rate only matter at the low SSFR end.  While our
low-metallicity outliers do have higher-than-normal SSFRs, this relative
difference alone is not enough to explain their low oxygen abundances.

\begin{figure}
\plotone{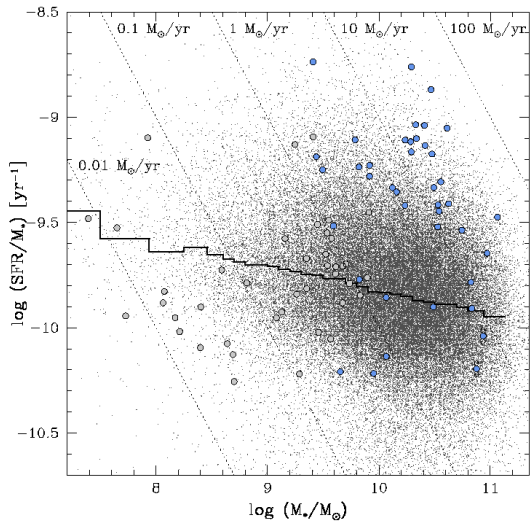}
\caption{\label{fig:sfrmm} Specific star formation rate vs.\ stellar
  mass.  The grey points are galaxies from the main \citet{tremonti04}
  SDSS sample described in \S\,\ref{sec:sample}, the low-metallicity
  massive galaxies in our sample are denoted with blue circles (with the
  only two clearly spiral galaxies marked by ``\S'' symbols), and the
  metal-rich dwarf galaxies from Paper~I are shown with large grey
  points for reference.  The histogram denotes the median specific star
  formation rate as a function of stellar mass and the dotted lines show
  lines of constant star formation rate.  The star formation rates
  estimates are from \citep{brinchmann04} and the stellar mass estimates
  are from \citet{tremonti04} and \citet{kauffmann03}; both have been
  aperture corrected. }
\end{figure}

\begin{figure*}
\includegraphics[height=\textwidth,angle=270]{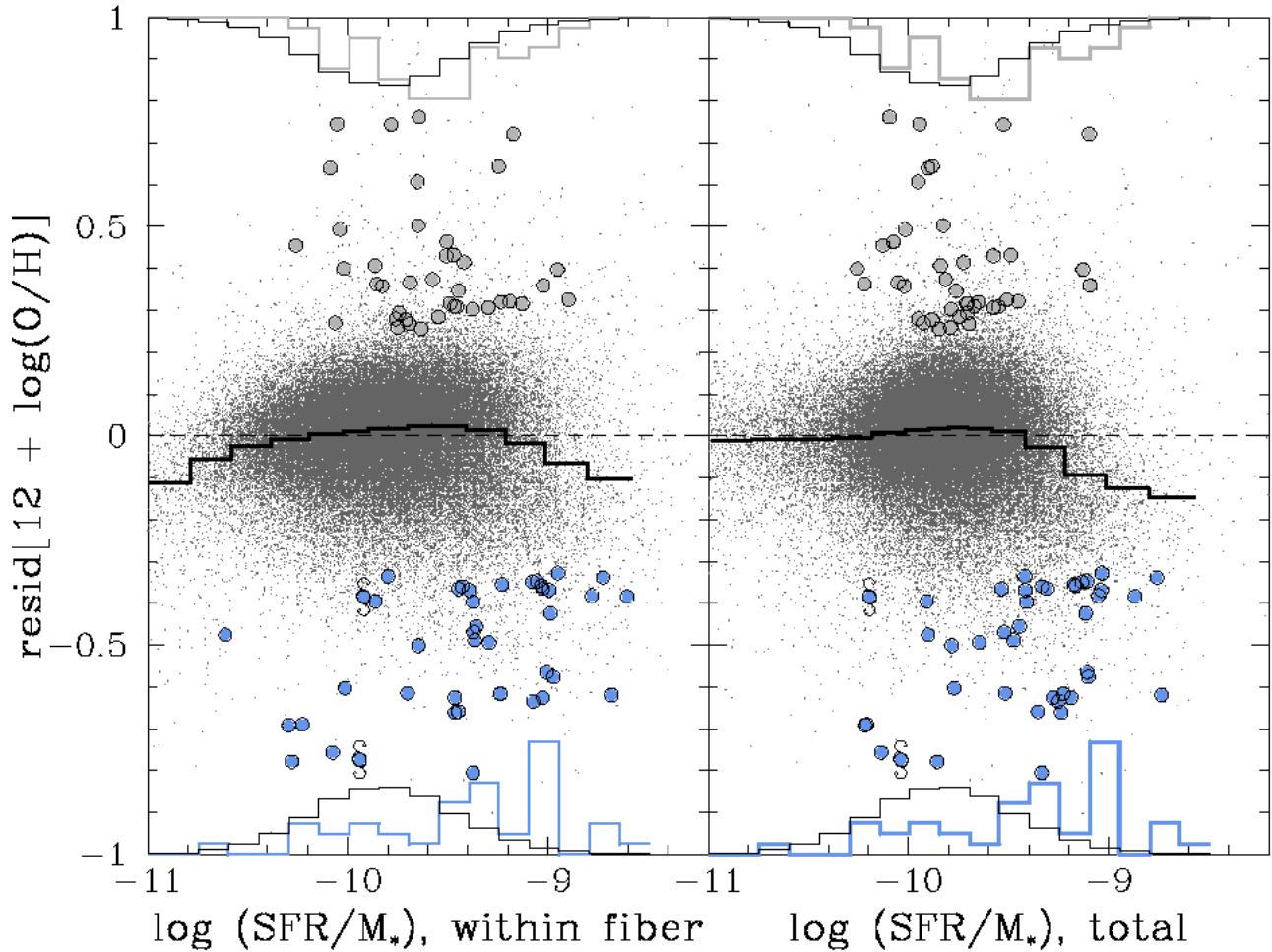}
\caption{\label{fig:sfroh} Residual in \tlogoh\ from \citet{tremonti04}
  fit to the mass--metallicity relation v.\ specific star formation rate
  (SSFR), both within the SDSS 3\arcsec\ fiber ({\em left}) and
  aperture-corrected ({\em right}). In each panel, the blue points are
  the high-mass low-metallicity outliers in our sample while the large
  grey points are the metal-rich dwarf galaxies from Paper~I. (The
  ``\S'' symbols denote the only two clearly spiral galaxies in the
  low-metallicity high-mass sample.) The grey cloud of points are the
  galaxies from the main \citet{tremonti04} SDSS sample described in
  \S\,\ref{sec:sample}, and the central thick black histogram is the
  median residual \tlogoh\ in bins of SSFR.  The thin black histograms
  at the top and bottom of each panel are the distribution of SSFR of
  the parent sample, while the blue histograms at the bottom of each
  panel is the SSFR distribution of the low-metallicity outliers and the
  grey histogram at the top of each panel is the SSFR distribution of
  the high-metallicity outliers.}
\end{figure*}

\subsection{Observed Metallicity of Merging Galaxies}\label{sec:merge}
As can be seen in Figure~\ref{fig:images}, most of galaxies in
this sample are tidally interacting.  Simulations have shown that galaxy
major mergers are accompanied by starbursts and gas inflow from large
galactocentric radii \citep[e.g.,][]{barnes92, mihos96,cox06}.  Because
typical spiral galaxies have metallicity gradients \citep{kennicutt03},
the inflowing gas will have a lower oxygen abundance than the native
central gas.  This suggests that perhaps our galaxies' observed low
metallicities are a consequence of large quantities of low-metallicity
gas from large galactocentric radii are being transported into the
central few kiloparsecs and diluting the metal-rich central gas.

We can test this idea at the order of magnitude level to verify that it
can give the observed oxygen abundance offsets of $-0.3$ to $-0.8$\,dex
(see, e.g., Fig.\ \ref{fig:ohff}).  The SDSS data measure the galaxies'
{\em central} oxygen abundances; we wish to know what the change in
oxygen abundance would be if low-metallicity gas from larger radii were
dumped into the central regions.  If we assume that {\em all} of the gas
out to some large radius $R$ was to participate in this inflow, then the
lower abundance would be the same as if we simply measured the total
oxygen abundance for the entire galaxy gas disk out to the radius $R$.
For reference, typical observed \ion{H}{1} disks (e.g.,
\citealt{boomsma08} in NGC~6946) have radii of $R\sim 20$\,kpc.  We
consider a model spiral galaxy with a gradient of \tlogoh\ of slope
$\Gamma_{\mbox{\scriptsize O/H}}$ in units of dex per kiloparsec;
$\Gamma_{\mbox{\scriptsize O/H}}$ is and is typically of order
$-0.05$\,dex\,kpc$^{-1}$ \citep{zaritsky94,vanzee98,kennicutt03}.  We
willl also assume that our model galaxy has a gas surface density
$\Sigma$ that obeys a power-law relation with the radius such that
$\Sigma(R) = \Sigma_0 (R/R_0)^{\alpha}$.  While a constant surface
density ($\alpha=0$) is a reasonable assumption at large radii
\citep{wong02,leroy08}, we will show that the calculated oxygen abundance
dilution is not strongly dependent on the choice of this power-law
slope.  For notational simplicity, in this section only we will use O/H
as shorthand for the more cumbersome \tlogoh.

Let $\langle\mbox{O/H}\rangle_{\mbox{\scriptsize SDSS}}$ be the mean O/H
within an SDSS spectroscopic fiber of a spiral galaxy (i.e., within the
central 3\arcsec\ diameter).  Formally,
\begin{equation}\label{eqn:s}
\langle\mbox{O/H}\rangle_{\mbox{\scriptsize SDSS}}\equiv \frac{\int\limits^{R_3}_{0} \Sigma(R)
  (\mbox{O}/\mbox{H})R\mbox{d}R\mbox{d}\theta}{\int\limits^{R_3}_{0} \Sigma(R) R\mbox{d}R\mbox{d}\theta},
\end{equation}
where $R_3$ is the radius in kiloparsecs of the galaxy at a radius of
1.5\arcsec\ (i.e., spanned by the 3\arcsec\ SDSS fiber; this corresponds
to $\approx 4.9$\,kpc at $z=0.2$) and we are implicitly assuming the
scale-height of the galaxy is roughly constant with radius.  For radii
$R>R_3$, the oxygen abundance is $\mbox{O/H} = (\mbox{O/H})_0 +
\Gamma_{\mbox{\scriptsize O/H}}R$, where $(\mbox{O/H})_0$ is the oxygen
abundance extrapolated to $R=0$ using the adopted the abundance
gradient. We can now expand Equation~(\ref{eqn:s}) to find
\begin{eqnarray}
\langle\mbox{O/H}\rangle_{\mbox{\scriptsize SDSS}} & =& \left[\frac{2+\alpha}{2\pi R_3^{2+\alpha}}\right] 2\pi
\int\limits^{R_3}_{0} R^{\alpha}[(\mbox{O/H})_0 +
\Gamma_{\mbox{\scriptsize O/H}}R]R\mbox{d}R \nonumber \\
& = &(\mbox{O/H})_0 + \left[\frac{2+\alpha}{3+\alpha}\right]\Gamma_{\mbox{\scriptsize O/H}}R_3.
\end{eqnarray}
Solving for the $R=0$ abundance, we find $(\mbox{O/H})_0 =
\langle\mbox{O/H}\rangle_{\mbox{\scriptsize
SDSS}}-[(2+\alpha)/(3+\alpha)]\Gamma_{\mbox{\scriptsize O/H}}R_3$.  The
mean metallicity $\langle \mbox{O/H}\rangle_{\leq R}$ out to a radius
$R$ is therefore
\begin{equation}
\langle \mbox{O/H}\rangle_{\leq R} = \frac{\int\limits^{R}_{0} \Sigma(R)
  (\mbox{O}/\mbox{H})R\mbox{d}R\mbox{d}\theta}{\int\limits^{R}_{0}
  \Sigma(R) R\mbox{d}R\mbox{d}\theta} = (\mbox{O/H})_0 + \left[\frac{2+\alpha}{3+\alpha}\right]\Gamma_{\mbox{\scriptsize O/H}}R.
\end{equation}
Hence the change in metallicity $\Delta(\mbox{O/H})$ we would expect
within an SDSS fiber diameter is $\Delta(\mbox{O/H}) =
[(2+\alpha)/(3+\alpha)]\Gamma_{\mbox{\scriptsize O/H}}\Delta R =
[(2+\alpha)/(3+\alpha)]\Gamma_{\mbox{\scriptsize O/H}}(R - R_3)$.  For
an outer radius of $R = 20$\,kpc, an inner radius of $R_3 = 5$\,kpc, an
abundance gradient of $\Gamma_{\mbox{\scriptsize O/H}} =
-0.05$\,dex\,kpc$^{-1}$, and a flat radial gas-surface density
distribution of $\alpha = 0$, this leads to an oxygen abundance dilution
of $\Delta(\mbox{O/H}) = -0.5$\,dex, which is in the range we observe
for these low-metallicity outliers (see, e.g., Fig.~\ref{fig:sfroh}).

We note that there are several caveats with this picture.  First, we
have assumed that the observed oxygen abundance corresponds to a
mass-weighted abundance, whereas because we measure the metallicity via
emission line spectra, it is actually weighted by the luminosities of
the \ion{H}{2} regions.  Secondly, it is unclear if the timescales
involved pose any problem.  Simulations show mergers {\em do} induce gas
inflow to the centers of galaxies \citep{cox06}, but most of this inflow
occurs during the first close encounter between the two galaxies.  While
a few of the galaxies in our sample do have nearby neighbors, most do
not: SDSS detects only a single, if disturbed, galaxy.  It is possible
that these are not yet dynamically relaxed; the two galaxies can pass
close enough to one another that they are seen by SDSS as one object.
If a merger-induced starburst lasts $\sim 1$\,Gyr \citep{cox06}, then
$\sim 1$\% of all (properly defined) starburst mergers should display
lower-than-expected metallicities.  The relevant dynamical timescale for
gas inflow depends on the typical radius from which the gas originates.
Because the dynamical time $t_{\mbox{\scriptsize dynam}} \approx
R/v_{\mbox{\scriptsize esc}}$, gas at larger radii takes longer to reach
$R_3$ and help dilute the central metallicity.

Several studies have shown that galaxies in close pairs tend to have
lower metallicies than expected. \citet{kewley06a} found that galaxies
separated by $\lesssim 20 h^{-1}$\,kpc have systematically lower oxygen
abundances by $\sim 0.2$\,dex.  On the other hand, \citet{ellison08b}
found an offset of $\sim 0.05$\,dex in \tlogoh\ at fixed mass galaxies
in pairs with similar separations; the investigations of
\citet{michel08} of galaxies in close pairs suggest that the relative
\logoh\ offset is greater for the lower-mass galaxy in the pair than for
the higher-mass galaxy.  \citeauthor{kewley06a}\ also found relatively
high specific star formation rates for the lower-metallicity member in
their galaxy pairs; they attribute both the strong starburst activity
and the lower oxygen abundance to fresh gas inflowing to the galaxies'
central regions due to interaction with the closely neighboring galaxy.

In a similar vein, \citet{lee04} found that morphologically disturbed
galaxies have systematically lower oxygen abundances (or higher
luminosities) than galaxies of similar brightness (or metallicity).
Likewise, luminous infrared galaxies (LIRGs), whose high specific star
formation rates and highly disturbed morphologies are believed to be due
to a recent major merger, are observed to have oxygen abundances that
are $\sim 0.4$\,dex lower than other galaxies of comparable masses
\citep{rupke08}.  The pronounced blue colors of our outliers (see
Fig.~\ref{fig:cmd}) implies that while the star formation rates for many
of these galaxies are consistent with those of LIRGs, these outliers are
clearly not as dusty as their infrared-luminous cousins; it is unclear
the extent to which these classes of galaxies are related, and why, if
LIRGs do typically have such relatively low metallicities, no galaxies
in our sample are extremely red.

\section{Conclucions and Implications}\label{sec:conc}
We have identified a sample of 42 low-metallicity high-mass outliers
from the mass--metallicity relation.  As a population, these galaxies
have disturbed morphologies and high specific star formation rates,
implying that they are undergoing a merger-induced starburst.  We
propose that their observed low oxygen abundances are due the tidal
interaction inducing large-scale gas inflow which subsequently dilutes
the central interstellar medium in these galaxies.  While there have
been several observational and theoretical suggestions that interacting
galaxies will result in a decreased observed gas-phase metallicity, this
is the first work that has shown that the vast majority of severe
low-metallicity outliers from the mass--metallicity relation are
morphologically disturbed.  Finally, we note that the cuts outlined in
\S\,\ref{sec:sel} provide an effective means of using {\em only} colors
and spectra to identify a rather pure (though not necessarily complete)
sample of tidally disturbed galaxies.

One striking implication of these results is that, while it is safe to
assume that the metallicities for {\em populations} of galaxies will
fall within a particular range of values given their luminosities at
redshifts for which the luminosity--metallicity relation has been
measured, one should not assign metallicities to {\em individual}
galaxies based solely on their luminosities.  For example, while studies
of the host galaxies of long gamma-ray bursts (GRBs) suggest that GRBs
are only found in low-metallicity environments
\citep{stanek06,kewley07}, several recent GRBs have been associated with
very luminous hosts, such as GRB~070306 and its $M_B \sim -22.3$ host
galaxy \citep{jaunsen08}.  As we have shown in this paper and Paper~I,
assuming an oxygen abundance for a galaxy from its luminosity alone can
result in mis-estimating \tlogoh\ in luminous galaxies by a whole dex,
and thus one should not assume that, e.g., the host of GRB~070306
necessarily has a high metallicity.  For cases in which \logoh\ has been
measured via emission line spectra for GRB hosts, it is generally low,
even if the galaxy is luminous.  For example, the host galaxy of
GRB~031203 has an absolute $B$-band magnitude similar to that of the
Milky Way, yet \citet{margutti07} find it to have a low metallicity
($12+\log[\mbox{O}/\mbox{H}] = 8.12$) as well as a high star formation
rate ($\sim 13\, \mbox{M}_{\odot}$\,yr$^{-1}$).  \citet{prochaska04}
interpret this offset in metallicity as a sign that GRB~031203 went off
in a ``very young star-forming region,'' which is in line with our
interpretation in \S\,\ref{sec:merge}.  Futhermore, the brighter GRB
hosts studied by \citet{fruchter06} are morphologically very similar to
our massive low-metallicity galaxies, and our results strongly suggest
that the oxygen abundances inferred from luminosity alone are uncertain
by as much as 1\,dex. Specifically, if a luminous galaxy is
morphologically disturbed, has a high specific star formation rate, and
is extremely blue, then it should not be assumed to lie within the
luminosity--metallicity locus of star-forming galaxies.

\acknowledgements{ We would like to thank Roberto Assef for providing us
   with access to his low-resolution templates and codes for generating
   Figure~\ref{fig:cmd} and Jason Eastman for assistance with making
   Figure~\ref{fig:images}.  We are also grateful to David Weinberg for
   comments on the text as well as Todd Thompson, Paul Martini, and
   Scott Gaudi for helpful discussions.  We further thank Sara Ellison,
   Maryam Modjaz, Janice Lee, Curt Struck, and Warren Brown for useful
   feedback on the first draft of this paper, as well as Roderik
   Overzier for pointing out to us that one of our outliers was in the
   \citet{overzier08} sample.  Finally, we thank the anonymous referee
   for their thoughtful suggestions.

    Funding for the SDSS and SDSS-II has been provided by the Alfred
    P.\ Sloan Foundation, the Participating Institutions, the National
    Science Foundation, the U.S. Department of Energy, the National
    Aeronautics and Space Administration, the Japanese Monbukagakusho,
    the Max Planck Society, and the Higher Education Funding Council for
    England. The SDSS Web Site is \texttt{http://www.sdss.org/}.

    The SDSS is managed by the Astrophysical Research Consortium for the
    Participating Institutions. The Participating Institutions are the
    American Museum of Natural History, Astrophysical Institute Potsdam,
    University of Basel, University of Cambridge, Case Western Reserve
    University, University of Chicago, Drexel University, Fermilab, the
    Institute for Advanced Study, the Japan Participation Group, Johns
    Hopkins University, the Joint Institute for Nuclear Astrophysics,
    the Kavli Institute for Particle Astrophysics and Cosmology, the
    Korean Scientist Group, the Chinese Academy of Sciences (LAMOST),
    Los Alamos National Laboratory, the Max-Planck-Institute for
    Astronomy (MPIA), the Max-Planck-Institute for Astrophysics (MPA),
    New Mexico State University, Ohio State University, University of
    Pittsburgh, University of Portsmouth, Princeton University, the
    United States Naval Observatory, and the University of Washington.

The STARLIGHT project is supported by the Brazilian agencies CNPq, CAPES
and FAPESP and by the France-Brazil CAPES/Cofecub }


\begin{thebibliography}{47}
\expandafter\ifx\csname natexlab\endcsname\relax\def\natexlab#1{#1}\fi

\bibitem[{{Adelman-McCarthy} {et~al.}(2008){Adelman-McCarthy}, {Ag{\"u}eros},
  {Allam}, {Allende Prieto}, {Anderson}, {Anderson}, {Annis}, {Bahcall},
  {Bailer-Jones}, {Baldry}, {Barentine}, {Bassett}, {Becker}, {Beers}, {Bell},
  {Berlind}, {Bernardi}, {Blanton}, {Bochanski}, {Boroski}, {Brinchmann},
  {Brinkmann}, {Brunner}, {Budav{\'a}ri}, {Carliles}, {Carr}, {Castander},
  {Cinabro}, {Cool}, {Covey}, {Csabai}, {Cunha}, {Davenport}, {Dilday}, {Doi},
  {Eisenstein}, {Evans}, {Fan}, {Finkbeiner}, {Friedman}, {Frieman},
  {Fukugita}, {G{\"a}nsicke}, {Gates}, {Gillespie}, {Glazebrook}, {Gray},
  {Grebel}, {Gunn}, {Gurbani}, {Hall}, {Harding}, {Harvanek}, {Hawley},
  {Hayes}, {Heckman}, {Hendry}, {Hindsley}, {Hirata}, {Hogan}, {Hogg}, {Hyde},
  {Ichikawa}, {Ivezi{\'c}}, {Jester}, {Johnson}, {Jorgensen}, {Juri{\'c}},
  {Kent}, {Kessler}, {Kleinman}, {Knapp}, {Kron}, {Krzesinski}, {Kuropatkin},
  {Lamb}, {Lampeitl}, {Lebedeva}, {Lee}, {Leger}, {L{\'e}pine}, {Lima}, {Lin},
  {Long}, {Loomis}, {Loveday}, {Lupton}, {Malanushenko}, {Malanushenko},
  {Mandelbaum}, {Margon}, {Marriner}, {Mart{\'{\i}}nez-Delgado}, {Matsubara},
  {McGehee}, {McKay}, {Meiksin}, {Morrison}, {Munn}, {Nakajima}, {Neilsen},
  {Newberg}, {Nichol}, {Nicinski}, {Nieto-Santisteban}, {Nitta}, {Okamura},
  {Owen}, {Oyaizu}, {Padmanabhan}, {Pan}, {Park}, {Peoples}, {Pier}, {Pope},
  {Purger}, {Raddick}, {Re Fiorentin}, {Richards}, {Richmond}, {Riess}, {Rix},
  {Rockosi}, {Sako}, {Schlegel}, {Schneider}, {Schreiber}, {Schwope}, {Seljak},
  {Sesar}, {Sheldon}, {Shimasaku}, {Sivarani}, {Smith}, {Snedden}, {Steinmetz},
  {Strauss}, {SubbaRao}, {Suto}, {Szalay}, {Szapudi}, {Szkody}, {Tegmark},
  {Thakar}, {Tremonti}, {Tucker}, {Uomoto}, {Vanden Berk}, {Vandenberg},
  {Vidrih}, {Vogeley}, {Voges}, {Vogt}, {Wadadekar}, {Weinberg}, {West},
  {White}, {Wilhite}, {Yanny}, {Yocum}, {York}, {Zehavi}, \&
  {Zucker}}]{adelman08}
{Adelman-McCarthy}, J.~K., et al. 2008, \apjs, 175, 297

\bibitem[{{Adelman-McCarthy} {et~al.}(2006){Adelman-McCarthy}, {Ag{\"u}eros},
  {Allam}, {Anderson}, {Anderson}, {Annis}, {Bahcall}, {Baldry}, {Barentine},
  {Berlind}, {Bernardi}, {Blanton}, {Boroski}, {Brewington}, {Brinchmann},
  {Brinkmann}, {Brunner}, {Budav{\'a}ri}, {Carey}, {Carr}, {Castander},
  {Connolly}, {Csabai}, {Czarapata}, {Dalcanton}, {Doi}, {Dong}, {Eisenstein},
  {Evans}, {Fan}, {Finkbeiner}, {Friedman}, {Frieman}, {Fukugita}, {Gillespie},
  {Glazebrook}, {Gray}, {Grebel}, {Gunn}, {Gurbani}, {de Haas}, {Hall},
  {Harris}, {Harvanek}, {Hawley}, {Hayes}, {Hendry}, {Hennessy}, {Hindsley},
  {Hirata}, {Hogan}, {Hogg}, {Holmgren}, {Holtzman}, {Ichikawa}, {Ivezi{\'c}},
  {Jester}, {Johnston}, {Jorgensen}, {Juri{\'c}}, {Kent}, {Kleinman}, {Knapp},
  {Kniazev}, {Kron}, {Krzesinski}, {Kuropatkin}, {Lamb}, {Lampeitl}, {Lee},
  {Leger}, {Lin}, {Long}, {Loveday}, {Lupton}, {Margon},
  {Mart{\'{\i}}nez-Delgado}, {Mandelbaum}, {Matsubara}, {McGehee}, {McKay},
  {Meiksin}, {Munn}, {Nakajima}, {Nash}, {Neilsen}, {Newberg}, {Newman},
  {Nichol}, {Nicinski}, {Nieto-Santisteban}, {Nitta}, {O'Mullane}, {Okamura},
  {Owen}, {Padmanabhan}, {Pauls}, {Peoples}, {Pier}, {Pope}, {Pourbaix},
  {Quinn}, {Richards}, {Richmond}, {Rockosi}, {Schlegel}, {Schneider},
  {Schroeder}, {Scranton}, {Seljak}, {Sheldon}, {Shimasaku}, {Smith}, {Smol{\v
  c}i{\'c}}, {Snedden}, {Stoughton}, {Strauss}, {SubbaRao}, {Szalay},
  {Szapudi}, {Szkody}, {Tegmark}, {Thakar}, {Tucker}, {Uomoto}, {Vanden Berk},
  {Vandenberg}, {Vogeley}, {Voges}, {Vogt}, {Walkowicz}, {Weinberg}, {West},
  {White}, {Xu}, {Yanny}, {Yocum}, {York}, {Zehavi}, {Zibetti}, \&
  {Zucker}}]{adelman06}
{Adelman-McCarthy}, J.~K., et al. 2006, \apjs, 162, 38

\bibitem[{{Assef} {et~al.}(2008){Assef}, {Kochanek}, {Brodwin}, {Brown},
  {Caldwell}, {Cool}, {Eisenhardt}, {Eisenstein}, {Gonzalez}, {Jannuzi},
  {Jones}, {McKenzie}, {Murray}, \& {Stern}}]{assef08}
{Assef}, R.~J., {Kochanek}, C.~S., {Brodwin}, M., {Brown}, M.~J.~I.,
  {Caldwell}, N., {Cool}, R.~J., {Eisenhardt}, P., {Eisenstein}, D.,
  {Gonzalez}, A.~H., {Jannuzi}, B.~T., {Jones}, C., {McKenzie}, E., {Murray},
  S.~S., \& {Stern}, D. 2008, \apj, 676, 286

\bibitem[{{Baldwin} {et~al.}(1981){Baldwin}, {Phillips}, \&
  {Terlevich}}]{baldwin81}
{Baldwin}, J.~A., {Phillips}, M.~M., \& {Terlevich}, R. 1981, \pasp, 93, 5

\bibitem[{{Barnes} \& {Hernquist}(1992)}]{barnes92}
{Barnes}, J.~E. \& {Hernquist}, L. 1992, \araa, 30, 705

\bibitem[{{Boomsma} {et~al.}(2008){Boomsma}, {Oosterloo}, {Fraternali}, {van
  der Hulst}, \& {Sancisi}}]{boomsma08}
{Boomsma}, R., {Oosterloo}, T.~A., {Fraternali}, F., {van der Hulst}, J.~M., \&
  {Sancisi}, R. 2008, ArXiv e-prints, arXiv:0807.3339

\bibitem[{{Bresolin}(2006)}]{bresolin06}
{Bresolin}, F. 2006, ArXiv Astrophysics e-prints, astro-ph/0610690

\bibitem[{{Brinchmann} {et~al.}(2004){Brinchmann}, {Charlot}, {White},
  {Tremonti}, {Kauffmann}, {Heckman}, \& {Brinkmann}}]{brinchmann04}
{Brinchmann}, J., {Charlot}, S., {White}, S.~D.~M., {Tremonti}, C.,
  {Kauffmann}, G., {Heckman}, T., \& {Brinkmann}, J. 2004, \mnras, 351, 1151

\bibitem[{{Cardelli} {et~al.}(1989){Cardelli}, {Clayton}, \&
  {Mathis}}]{cardelli89}
{Cardelli}, J.~A., {Clayton}, G.~C., \& {Mathis}, J.~S. 1989, \apj, 345, 245

\bibitem[{{Cid Fernandes} {et~al.}(2005){Cid Fernandes}, {Mateus}, {Sodr{\'e}},
  {Stasi{\'n}ska}, \& {Gomes}}]{fernandes05}
{Cid Fernandes}, R., {Mateus}, A., {Sodr{\'e}}, L., {Stasi{\'n}ska}, G., \&
  {Gomes}, J.~M. 2005, \mnras, 358, 363

\bibitem[{{Cox} {et~al.}(2006){Cox}, {Jonsson}, {Primack}, \&
  {Somerville}}]{cox06}
{Cox}, T.~J., {Jonsson}, P., {Primack}, J.~R., \& {Somerville}, R.~S. 2006,
  \mnras, 373, 1013

\bibitem[{{Dalcanton}(2007)}]{dalcanton07}
{Dalcanton}, J.~J. 2007, \apj, 658, 941

\bibitem[{{Ellison} {et~al.}(2008{\natexlab{a}}){Ellison}, {Patton}, {Simard},
  \& {McConnachie}}]{ellison08a}
{Ellison}, S.~L., {Patton}, D.~R., {Simard}, L., \& {McConnachie}, A.~W.
  2008{\natexlab{a}}, \apjl, 672, L107

\bibitem[{{Ellison} {et~al.}(2008{\natexlab{b}}){Ellison}, {Patton}, {Simard},
  \& {McConnachie}}]{ellison08b}
---. 2008{\natexlab{b}}, \aj, 135, 1877

\bibitem[{{Erb} {et~al.}(2006){Erb}, {Shapley}, {Pettini}, {Steidel}, {Reddy},
  \& {Adelberger}}]{erb06a}
{Erb}, D.~K., {Shapley}, A.~E., {Pettini}, M., {Steidel}, C.~C., {Reddy},
  N.~A., \& {Adelberger}, K.~L. 2006, \apj, 644, 813

\bibitem[{{Finlator} \& {Dav{\'e}}(2008)}]{finlator08}
{Finlator}, K. \& {Dav{\'e}}, R. 2008, \mnras, 385, 2181

\bibitem[{{Fruchter} {et~al.}(2006){Fruchter}, {Levan}, {Strolger},
  {Vreeswijk}, {Thorsett}, {Bersier}, {Burud}, {Castro Cer{\'o}n},
  {Castro-Tirado}, {Conselice}, {Dahlen}, {Ferguson}, {Fynbo}, {Garnavich},
  {Gibbons}, {Gorosabel}, {Gull}, {Hjorth}, {Holland}, {Kouveliotou}, {Levay},
  {Livio}, {Metzger}, {Nugent}, {Petro}, {Pian}, {Rhoads}, {Riess}, {Sahu},
  {Smette}, {Tanvir}, {Wijers}, \& {Woosley}}]{fruchter06}
{Fruchter}, A.~S., {Levan}, A.~J., {Strolger}, L., {Vreeswijk}, P.~M.,
  {Thorsett}, S.~E., {Bersier}, D., {Burud}, I., {Castro Cer{\'o}n}, J.~M.,
  {Castro-Tirado}, A.~J., {Conselice}, C., {Dahlen}, T., {Ferguson}, H.~C.,
  {Fynbo}, J.~P.~U., {Garnavich}, P.~M., {Gibbons}, R.~A., {Gorosabel}, J.,
  {Gull}, T.~R., {Hjorth}, J., {Holland}, S.~T., {Kouveliotou}, C., {Levay},
  Z., {Livio}, M., {Metzger}, M.~R., {Nugent}, P.~E., {Petro}, L., {Pian}, E.,
  {Rhoads}, J.~E., {Riess}, A.~G., {Sahu}, K.~C., {Smette}, A., {Tanvir},
  N.~R., {Wijers}, R.~A.~M.~J., \& {Woosley}, S.~E. 2006, \nat, 441, 463

\bibitem[{{Hoopes} {et~al.}(2007){Hoopes}, {Heckman}, {Salim}, {Seibert},
  {Tremonti}, {Schiminovich}, {Rich}, {Martin}, {Charlot}, {Kauffmann},
  {Forster}, {Friedman}, {Morrissey}, {Neff}, {Small}, {Wyder}, {Bianchi},
  {Donas}, {Lee}, {Madore}, {Milliard}, {Szalay}, {Welsh}, \& {Yi}}]{hoopes07}
{Hoopes}, C.~G., {Heckman}, T.~M., {Salim}, S., {Seibert}, M., {Tremonti},
  C.~A., {Schiminovich}, D., {Rich}, R.~M., {Martin}, D.~C., {Charlot}, S.,
  {Kauffmann}, G., {Forster}, K., {Friedman}, P.~G., {Morrissey}, P., {Neff},
  S.~G., {Small}, T., {Wyder}, T.~K., {Bianchi}, L., {Donas}, J., {Lee}, Y.-W.,
  {Madore}, B.~F., {Milliard}, B., {Szalay}, A.~S., {Welsh}, B.~Y., \& {Yi},
  S.~K. 2007, \apjs, 173, 441

\bibitem[{{Jaunsen} {et~al.}(2008){Jaunsen}, {Rol}, {Watson}, {Malesani},
  {Fynbo}, {Milvang-Jensen}, {Hjorth}, {Vreeswijk}, {Ovaldsen}, {Wiersema},
  {Tanvir}, {Gorosabel}, {Levan}, {Schirmer}, \& {Castro-Tirado}}]{jaunsen08}
{Jaunsen}, A.~O., {Rol}, E., {Watson}, D.~J., {Malesani}, D., {Fynbo},
  J.~P.~U., {Milvang-Jensen}, B., {Hjorth}, J., {Vreeswijk}, P.~M., {Ovaldsen},
  J.-E., {Wiersema}, K., {Tanvir}, N.~R., {Gorosabel}, J., {Levan}, A.~J.,
  {Schirmer}, M., \& {Castro-Tirado}, A.~J. 2008, \apj, 681, 453

\bibitem[{{Kauffmann} {et~al.}(2003){Kauffmann}, {Heckman}, {White}, {Charlot},
  {Tremonti}, {Brinchmann}, {Bruzual}, {Peng}, {Seibert}, {Bernardi},
  {Blanton}, {Brinkmann}, {Castander}, {Cs{\'a}bai}, {Fukugita}, {Ivezic},
  {Munn}, {Nichol}, {Padmanabhan}, {Thakar}, {Weinberg}, \&
  {York}}]{kauffmann03}
{Kauffmann}, G., {Heckman}, T.~M., {White}, S.~D.~M., {Charlot}, S.,
  {Tremonti}, C., {Brinchmann}, J., {Bruzual}, G., {Peng}, E.~W., {Seibert},
  M., {Bernardi}, M., {Blanton}, M., {Brinkmann}, J., {Castander}, F.,
  {Cs{\'a}bai}, I., {Fukugita}, M., {Ivezic}, Z., {Munn}, J.~A., {Nichol},
  R.~C., {Padmanabhan}, N., {Thakar}, A.~R., {Weinberg}, D.~H., \& {York}, D.
  2003, \mnras, 341, 33

\bibitem[{{Kennicutt} {et~al.}(2003){Kennicutt}, {Bresolin}, \&
  {Garnett}}]{kennicutt03}
{Kennicutt}, Jr., R.~C., {Bresolin}, F., \& {Garnett}, D.~R. 2003, \apj, 591,
  801

\bibitem[{{Kewley} {et~al.}(2007){Kewley}, {Brown}, {Geller}, {Kenyon}, \&
  {Kurtz}}]{kewley07}
{Kewley}, L.~J., {Brown}, W.~R., {Geller}, M.~J., {Kenyon}, S.~J., \& {Kurtz},
  M.~J. 2007, \aj, 133, 882

\bibitem[{{Kewley} \& {Dopita}(2002)}]{kewley02}
{Kewley}, L.~J. \& {Dopita}, M.~A. 2002, \apjs, 142, 35

\bibitem[{{Kewley} \& {Ellison}(2008)}]{kewley08}
{Kewley}, L.~J. \& {Ellison}, S.~L. 2008, \apj, 681, 1183

\bibitem[{{Kewley} {et~al.}(2006{\natexlab{a}}){Kewley}, {Geller}, \&
  {Barton}}]{kewley06a}
{Kewley}, L.~J., {Geller}, M.~J., \& {Barton}, E.~J. 2006{\natexlab{a}}, \aj,
  131, 2004

\bibitem[{{Kewley} {et~al.}(2006{\natexlab{b}}){Kewley}, {Groves}, {Kauffmann},
  \& {Heckman}}]{kewley06b}
{Kewley}, L.~J., {Groves}, B., {Kauffmann}, G., \& {Heckman}, T.
  2006{\natexlab{b}}, \mnras, 372, 961

\bibitem[{{Kobayashi} {et~al.}(2007){Kobayashi}, {Springel}, \&
  {White}}]{kobayashi07}
{Kobayashi}, C., {Springel}, V., \& {White}, S.~D.~M. 2007, \mnras, 376, 1465

\bibitem[{{Larson}(1974)}]{larson74}
{Larson}, R.~B. 1974, \mnras, 169, 229

\bibitem[{{Lee} {et~al.}(2004){Lee}, {Salzer}, \& {Melbourne}}]{lee04}
{Lee}, J.~C., {Salzer}, J.~J., \& {Melbourne}, J. 2004, \apj, 616, 752

\bibitem[{{Leroy} {et~al.}(2008){Leroy}, {Walter}, {Brinks}, {Bigiel}, {de
  Blok}, {Madore}, \& {Thornley}}]{leroy08}
{Leroy}, A.~K., {Walter}, F., {Brinks}, E., {Bigiel}, F., {de Blok}, W.~J.~G.,
  {Madore}, B., \& {Thornley}, M.~D. 2008, arXiv:0810.2556

\bibitem[{{Maiolino} {et~al.}(2008){Maiolino}, {Nagao}, {Grazian}, {Cocchia},
  {Marconi}, {Mannucci}, {Cimatti}, {Pipino}, {Ballero}, {Calura}, {Chiappini},
  {Fontana}, {Granato}, {Matteucci}, {Pastorini}, {Pentericci}, {Risaliti},
  {Salvati}, \& {Silva}}]{maiolino08}
{Maiolino}, R., {Nagao}, T., {Grazian}, A., {Cocchia}, F., {Marconi}, A.,
  {Mannucci}, F., {Cimatti}, A., {Pipino}, A., {Ballero}, S., {Calura}, F.,
  {Chiappini}, C., {Fontana}, A., {Granato}, G.~L., {Matteucci}, F.,
  {Pastorini}, G., {Pentericci}, L., {Risaliti}, G., {Salvati}, M., \& {Silva},
  L. 2008, \aap, 488, 463

\bibitem[{{Margutti} {et~al.}(2007){Margutti}, {Chincarini}, {Covino},
  {Tagliaferri}, {Campana}, {Della Valle}, {Filippenko}, {Fiore}, {Foley},
  {Fugazza}, {Giommi}, {Malesani}, {Moretti}, \& {Stella}}]{margutti07}
{Margutti}, R., {Chincarini}, G., {Covino}, S., {Tagliaferri}, G., {Campana},
  S., {Della Valle}, M., {Filippenko}, A.~V., {Fiore}, F., {Foley}, R.,
  {Fugazza}, D., {Giommi}, P., {Malesani}, D., {Moretti}, A., \& {Stella}, L.
  2007, \aap, 474, 815

\bibitem[{{McGaugh}(1991)}]{mcgaugh91}
{McGaugh}, S.~S. 1991, \apj, 380, 140

\bibitem[{{Michel-Dansac} {et~al.}(2008){Michel-Dansac}, {Lambas}, {Alonso}, \&
  {Tissera}}]{michel08}
{Michel-Dansac}, L., {Lambas}, D.~G., {Alonso}, M.~S., \& {Tissera}, P. 2008,
  \mnras, 386, L82

\bibitem[{{Mihos} \& {Hernquist}(1996)}]{mihos96}
{Mihos}, J.~C. \& {Hernquist}, L. 1996, \apj, 464, 641

\bibitem[{{Overzier} {et~al.}(2008){Overzier}, {Heckman}, {Kauffmann},
  {Seibert}, {Rich}, {Basu-Zych}, {Lotz}, {Aloisi}, {Charlot}, {Hoopes},
  {Martin}, {Schiminovich}, \& {Madore}}]{overzier08}
{Overzier}, R.~A., {Heckman}, T.~M., {Kauffmann}, G., {Seibert}, M., {Rich},
  R.~M., {Basu-Zych}, A., {Lotz}, J., {Aloisi}, A., {Charlot}, S., {Hoopes},
  C., {Martin}, D.~C., {Schiminovich}, D., \& {Madore}, B. 2008, \apj, 677, 37

\bibitem[{{Peeples} {et~al.}(2008){Peeples}, {Pogge}, \& {Stanek}}]{peeples08}
{Peeples}, M.~S., {Pogge}, R.~W., \& {Stanek}, K.~Z. 2008, \apj, 685, 904

\bibitem[{{Pettini} \& {Pagel}(2004)}]{pettini04}
{Pettini}, M. \& {Pagel}, B.~E.~J. 2004, \mnras, 348, L59

\bibitem[{{Prochaska} {et~al.}(2004){Prochaska}, {Bloom}, {Chen}, {Hurley},
  {Melbourne}, {Dressler}, {Graham}, {Osip}, \& {Vacca}}]{prochaska04}
{Prochaska}, J.~X., {Bloom}, J.~S., {Chen}, H.-W., {Hurley}, K.~C.,
  {Melbourne}, J., {Dressler}, A., {Graham}, J.~R., {Osip}, D.~J., \& {Vacca},
  W.~D. 2004, \apj, 611, 200

\bibitem[{{Rupke} {et~al.}(2008){Rupke}, {Veilleux}, \& {Baker}}]{rupke08}
{Rupke}, D.~S.~N., {Veilleux}, S., \& {Baker}, A.~J. 2008, \apj, 674, 172

\bibitem[{{Savaglio} {et~al.}(2005){Savaglio}, {Glazebrook}, {Le Borgne},
  {Juneau}, {Abraham}, {Chen}, {Crampton}, {McCarthy}, {Carlberg}, {Marzke},
  {Roth}, {J{\o}rgensen}, \& {Murowinski}}]{savaglio05}
{Savaglio}, S., {Glazebrook}, K., {Le Borgne}, D., {Juneau}, S., {Abraham},
  R.~G., {Chen}, H.-W., {Crampton}, D., {McCarthy}, P.~J., {Carlberg}, R.~G.,
  {Marzke}, R.~O., {Roth}, K., {J{\o}rgensen}, I., \& {Murowinski}, R. 2005,
  \apj, 635, 260

\bibitem[{{Stanek} {et~al.}(2006){Stanek}, {Gnedin}, {Beacom}, {Gould},
  {Johnson}, {Kollmeier}, {Modjaz}, {Pinsonneault}, {Pogge}, \&
  {Weinberg}}]{stanek06}
{Stanek}, K.~Z., {Gnedin}, O.~Y., {Beacom}, J.~F., {Gould}, A.~P., {Johnson},
  J.~A., {Kollmeier}, J.~A., {Modjaz}, M., {Pinsonneault}, M.~H., {Pogge}, R.,
  \& {Weinberg}, D.~H. 2006, Acta Astronomica, 56, 333

\bibitem[{{Struck} {et~al.}(2008){Struck}, {Hancock}, {Smith}, {Appleton},
  {Charmandaris}, \& {Giroux}}]{struck08}
{Struck}, C., {Hancock}, M., {Smith}, B.~J., {Appleton}, P.~N., {Charmandaris},
  V., \& {Giroux}, M. 2008, in IAU Symposium, Vol. 244, IAU Symposium, ed.
  J.~{Davies} \& M.~{Disney}, 231--234

\bibitem[{{Tremonti} {et~al.}(2004){Tremonti}, {Heckman}, {Kauffmann},
  {Brinchmann}, {Charlot}, {White}, {Seibert}, {Peng}, {Schlegel}, {Uomoto},
  {Fukugita}, \& {Brinkmann}}]{tremonti04}
{Tremonti}, C.~A., {Heckman}, T.~M., {Kauffmann}, G., {Brinchmann}, J.,
  {Charlot}, S., {White}, S.~D.~M., {Seibert}, M., {Peng}, E.~W., {Schlegel},
  D.~J., {Uomoto}, A., {Fukugita}, M., \& {Brinkmann}, J. 2004, \apj, 613, 898

\bibitem[{{van Zee} {et~al.}(1998){van Zee}, {Salzer}, {Haynes}, {O'Donoghue},
  \& {Balonek}}]{vanzee98}
{van Zee}, L., {Salzer}, J.~J., {Haynes}, M.~P., {O'Donoghue}, A.~A., \&
  {Balonek}, T.~J. 1998, \aj, 116, 2805

\bibitem[{{Wong} \& {Blitz}(2002)}]{wong02}
{Wong}, T. \& {Blitz}, L. 2002, \apj, 569, 157

\bibitem[{{Zaritsky} {et~al.}(1994){Zaritsky}, {Kennicutt}, \&
  {Huchra}}]{zaritsky94}
{Zaritsky}, D., {Kennicutt}, Jr., R.~C., \& {Huchra}, J.~P. 1994, \apj, 420, 87

\end{thebibliography}
\end{document}